\documentclass{aa}
\usepackage{epsfig}
\usepackage{amsmath}
\usepackage{txfonts}
\usepackage{natbib}
\usepackage{bm}

\begin{document}

\title{First direct observation of a torsional Alfv{\'e}n oscillation at coronal heights}

\author{P. Kohutova\inst{1,2}, E. Verwichte\inst{3} \and C. Froment\inst{1,2,4}}
\institute{Rosseland Centre for Solar Physics, University of Oslo, P.O. Box 1029, Blindern, NO-0315 Oslo, Norway\\
\email{petra.kohutova@astro.uio.no}
\and
Institute of Theoretical Astrophysics, University of Oslo, P.O. Box 1029, Blindern, NO-0315 Oslo, Norway
\and
Centre for Fusion, Space and Astrophysics, Department of Physics, University of Warwick, Coventry CV4 7AL, UK
\and 
LPC2E, CNRS and University of Orl\'eans, 3A avenue de la Recherche Scientifique, Orl\'eans, France \\}
\date{Received; accepted}

\abstract
{Torsional Alfv{\'e}n waves are promising candidates for transport of energy across different layers of the solar atmosphere and have been theoretically predicted for decades. Previous detections of Alfv{\'e}n waves so far have however mostly relied on indirect signatures.} 
{We present a first direct observational evidence of a fully resolved torsional Alfv{\'e}n oscillation of a large-scale structure occurring at coronal heights.}
{We analyse IRIS imaging and spectral observation of a surge resulting from magnetic reconnection between active region prominence threads and surrounding magnetic fieldlines.}
{The IRIS spectral data provides clear evidence of an oscillation in the line-of-sight velocity with a 180$^\circ$ phase difference between the oscillation signatures at opposite edges of the surge flux tube. This together with an alternating tilt in the Si IV and Mg II k spectra across the flux tube and the trajectories traced by the individual threads of the surge material provides clear evidence of torsional oscillation of the flux tube.}
{Our observation shows that magnetic reconnection leads to the generation of large-scale torsional Alfvén waves.}

\keywords{Magnetohydrodynamics (MHD) -- Sun: corona -- Sun: oscillations -- Sun: magnetic fields}

\maketitle

\section{Introduction}
The magnetized solar atmosphere supports a variety of magnetohydrodynamic waves \citep{nakariakov_coronal_2005}. In homogeneous plasma, such waves are either magnetoacoustic or purely magnetic. It should be noted that this distinction is only clear in specific geometries, as in realistic non-homogeneous plasma the different MHD modes are coupled and can have mixed properties \citep[e.g.][]{goossens_surface_2012}.  The magnetic, or Alfv{\'e}n waves are incompressible and dispersionless. They are difficult to dissipate in the absence of large gradients in local Alfv{\'e}n speed. They are therefore prime candidates for being responsible for transfer of energy between different layers of solar atmosphere.

In the case of an ideal magnetic flux tube which is a valid approximation for most magnetic structures observed in the solar atmosphere, the incompressible Alfvén mode corresponds to a torsional wave characterized by perturbations in the azimuthal components of velocity and magnetic field corresponding to a periodic axisymmetric rotation \citep{edwin_wave_1983}. Torsional Alfv{\'e}n wave excited along a coronal flux tube does not perturb the density of the flux tube and therefore does not modify the emission. Because of this, the torsional Alfvén modes are notoriously difficult to observe. The main observational signature of a torsional Alfv{\'e}n mode are opposite and alternating Doppler shifts at the opposite edges of the flux tube \citep{doorsselaere_detection_2008}. In practice, this corresponds to an alternating spectral line tilt seen in spectra taken across the oscillating structure.

\begin{figure*}
\includegraphics[width=43pc]{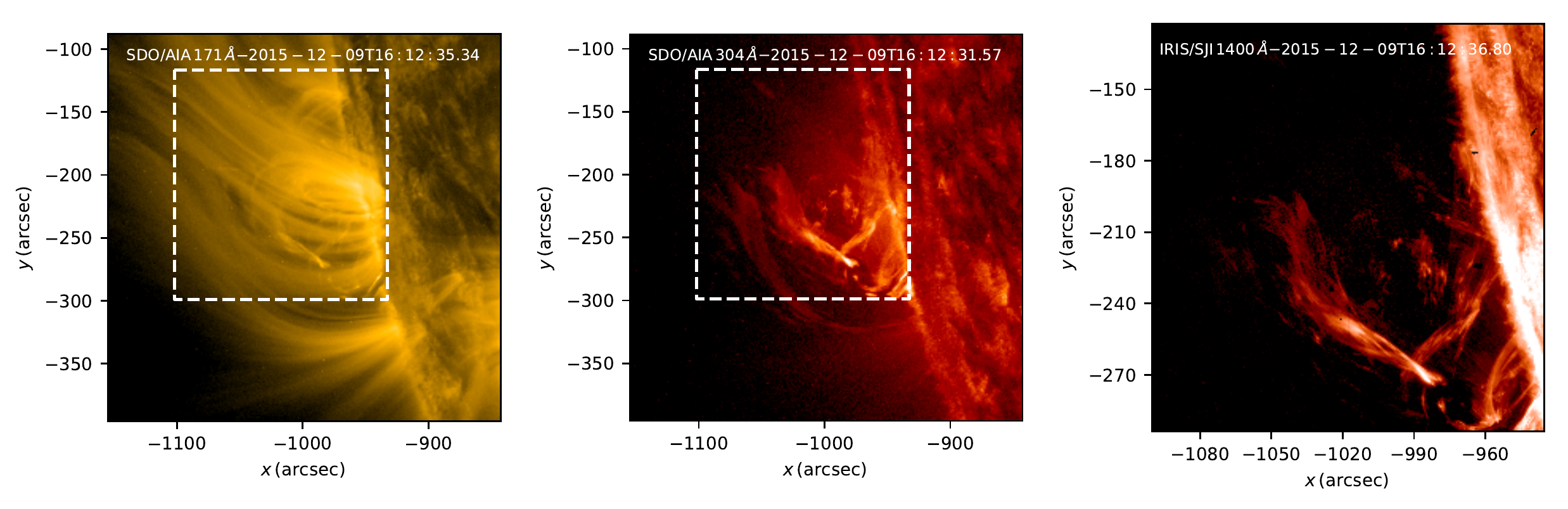}
\caption{Snapshots of the studied active region immediately following the surge eruption as observed at 16:12:35 in AIA 171 {\AA} (left), AIA 304 {\AA} (middle) and IRIS SJI 1400 {\AA} (right). The dashed line in the AIA images outlines the IRIS FOV. Animation of this figure is available.}
\label{fig:context}
\end{figure*}

Previous reports of Alfvén mode detection have so far mostly relied on indirect signatures, such as periodic broadening of spectral line width resulting from unresolved opposite Doppler shifts across the structure. Similar oscillations in the FWHM of the H$\alpha$ spectral line across magnetic bright point were reported by \citet{jess_alfven_2009}. Observations of opposite Doppler shifts in chromospheric spicules that are spatially resolved have been reported by \citet{srivastava_high_2017}, although showing only isolated cases of Doppler shift reversal and therefore lacking evidence of the oscillatory behaviour. Direct observational evidence of the torsional Alfvén mode in the solar atmosphere that is both spatially and temporally resolved is still missing.

Reports of uni-directional twisting motions are much more prevalent. Such motion has been observed in spicules \citep{pontieu_ubiquitous_2012}, swirls and tornados \citep{wedemeyer_magnetic_2012, shetye2019}, and during eruptive events consisting of ejection of solar plasma from the lower solar atmosphere into the corona as a result of magnetic reconnection, most notably in jets and surges \citep[e.g.][]{schmieder_twisting_2013, filippov_formation_2015,xue_observing_2016}. In that case, rotation is interpreted as an untwisting of the magnetic flux rope involved in the eruption \citep{iijima_three_2017}. The release of magnetic twist accumulated in the flux rope involved in the magnetic reconnection can result in overshooting the equilibrium of the newly formed flux tube and in the excitation of a torsional oscillation propagating at Alfvén speed. The mechanism of torsional wave generation by relaxation of twisted magnetic field has been proposed from theoretical models \citep{shibata_sweeping_1986, velli_alfven_1999, pariat_model_2009, torok_fanspine_2009}, but has not yet been observed. 

\section{Observation}

We analyse an event from 9 December 2015 occurring in in NOAA AR 12468 observed by the \textit{Interface Region Spectrograph} (IRIS) \citep{depontieu_interface_2014}. The IRIS level 2 slit-jaw imager (SJI) and spectrograph (SG) data used for analysis were taken between 16:12 and 17:11 UT. The SJI data are taken in two passbands; the far-UV (FUV) and the near-UV (NUV). The FUV passband is centred on 1400 {\AA} and is dominated by two Si IV lines formed at $\log T = 4.8$ in the transition region and the NUV passband is centred on 2796 {\AA} dominated by the Mg II K line core formed at $\log T = 4$ in the chromosphere, with an exposure time of 8 s, 19 s cadence and the image scale of $0.166 \arcsec$ pixel$^{-1}$. The IRIS observations were taken in sit-and-stare mode with a field-of-view of $167\arcsec \times 174\arcsec$ centered at $x,y = \left[-1017 \arcsec, -209 \arcsec \right]$. The spectrograph slit is fixed at $x = 1017\arcsec$.
 
We further use imaging EUV data from the \textit{Atmospheric Imaging Assembly} (AIA) on board the \textit{Solar Dynamics Observatory} (SDO) \citep{lemen_atmospheric_2012} in 171 {\AA} and 304 {\AA} passbands providing context images for plasma at coronal and transition region temperatures respectively. Level 1.5 SDO/AIA data have the image scale of $0.6 \arcsec$ pixel$^{-1}$, 12 s cadence and were normalised by the exposure time. 

We focus on the NOAA AR 12468 active region at the eastern limb (Fig.~\ref{fig:context}). Prominence with twisted flux-rope structure spanning across the whole active region can be observed at the limb in AIA 304 {\AA} and in the IRIS FUV and NUV SJI channels. The active region also contains footpoints of coronal loops with radii of the order of 100 Mm as well as several open magnetic field lines. Magnetic reconnection involving several prominence threads occurs in the foreground component of the prominence at 16:02 UT, when they can be observed to erupt and reconnect with surrounding open and closed magnetic field lines. The reconnection event is accompanied by a surge of cool plasma previously confined in the prominence flux rope. As the IRIS observing sequence starts at 16:12:36 UT, the analysis of the reconnection event itself is limited to imaging data from SDO/AIA. 

This IRIS dataset has been previously analysed by \citet{kohutova_formation_2019} and \citet{Shad2018} with a focus on coronal rain. In particular, \citet{kohutova_formation_2019} performed a detailed analysis of the reconnection event, the onset of thermal instability and associated coronal rain formation in one of the coronal loops. The analysis therein however excluded the upward-moving surge material and the associated downflows. In this study we focus on the evolution of the surge and the associated oscillatory motion following the reconnection.

\section{Torsional surge oscillation}

\begin{figure*}
\includegraphics[width=43pc]{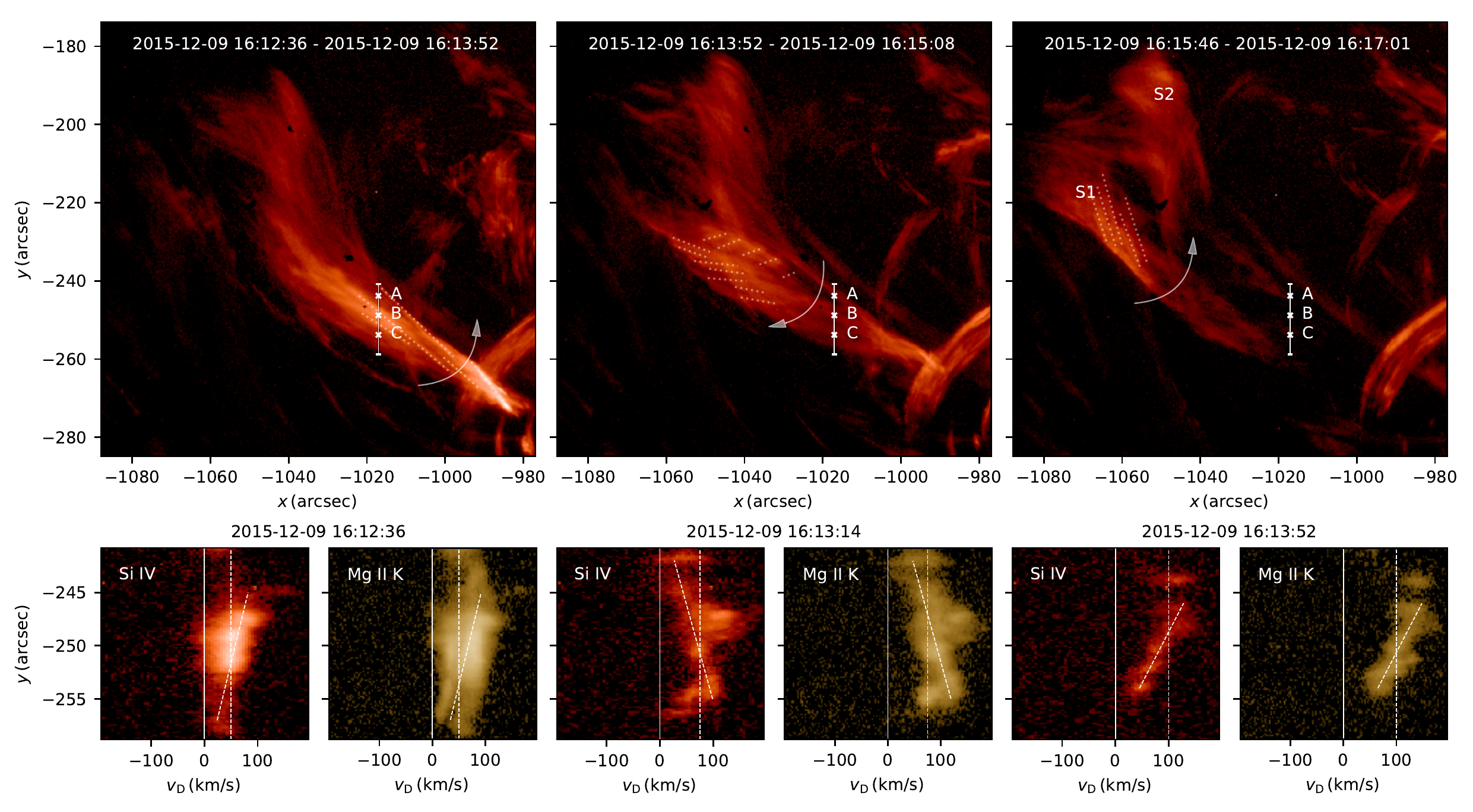}
\caption{Evolution of the rotating flux tube. Top: superimposed snapshots in IRIS FUV SJI data during three different phases of the torsional oscillation of the surge material. White dotted lines outline the helical trajectories of the plasma threads. The arrow shows the direction of the rotation. The white vertical line corresponds to the fraction of the slit intersecting the surge plasma. The freely moving and confined component of the ejected surge material are marked S1 and S2 respectively. Bottom: IRIS Si IV and Mg II K spectra of the surge material corresponding to the fraction of the slit marked by the white line. We show spectral snapshots from three different phases of the motion of the plasma at the slit position. Vertical dashed line marks the mean Doppler velocity of the plasma. The alternating tilt of the emission in the different phases, marked by tilted dashed line (obtained by visual examination) corresponds to the torsional motion of the surge material. Animation of this figure is available.}
\label{fig:flux_rotation}
\end{figure*}

\begin{figure*}
\includegraphics[width=43pc]{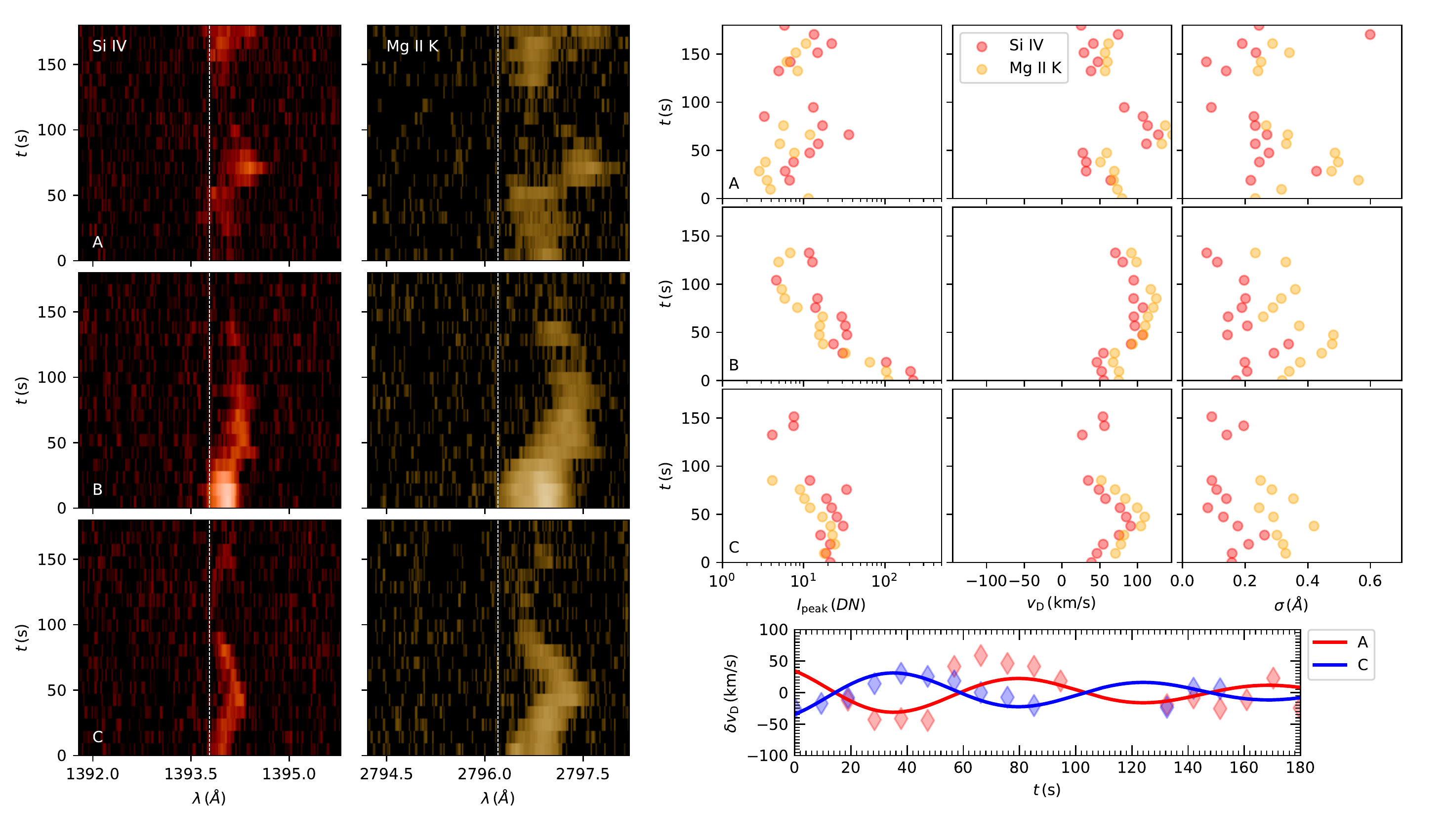}
\caption{Left: Evolution of Si IV and Mg II K line profiles during the initial surge phase at three different points A, B and C shown in Fig.~\ref{fig:flux_rotation} corresponding to the top edge of the surge, surge centre and bottom edge of the surge respectively. Left: Wavelength-time plot for Si IV (red) and Mg II K (orange) spectra during the initial 180 s when the surge material is detectable in the spectrograph data, starting at 16:12:36. Top right: Evolution of the parameters of the line profiles fitted with a single peak Gaussian. The scatter plots correspond to the peak intensity, Doppler velocity and line width at points A, B and C. We omit instances where the profiles could not be fitted reliably. Bottom right: Evolution of the line of sight velocities at the opposite edges of the flux tube fitted to detrended Si IV spectra. }
\label{fig:profile_evolution}
\end{figure*}

The reconnection of the individual prominence threads with surrounding coronal structures leads to a fraction of the cool prominence plasma being ejected away from the reconnection region. The bulk of the ejected material consists of two main components marked as S1 and S2 in Fig.~\ref{fig:flux_rotation}). Component S1 moves upwards freely and undergoes helical motion believed to be the result of the reconnection of a twisted bundle of prominence threads with overlying field lines. Helical motion of the ejected surge material suggests that the reconnection lead to the release of the magnetic twist accumulated in the the flux rope supporting the prominence. The ejection of component S2 is triggered by the reconnection of a prominence thread with a closed coronal loop. The confined material then moves along a closed loop-like trajectory and subsequently falls towards the solar surface.

The direction of the helical motion of the surge material alternates suggesting a large-scale torsional wave is set up in the flux tube following the reconnection. The helical motion of the individual threads is observable both in the imaging and spectral data. Rotation in clockwise direction viewed edge-on shows in spectral data as opposite red and blue-shifted emission on the top and bottom edges of the flux tube, relative to the mean line-of-sight velocity. The reverse is true for counter-clockwise rotation. The IRIS spectrograph slit is in sit-and-stare mode and crosses the surge material at an angle. As a result, it is possible to determine the presence and direction of rotation from variations of the Doppler shift as a function of distance along the slit. Figure \ref{fig:flux_rotation} shows SJI snapshots at various phases of the torsional motion of the flux tube as well as the Si IV and Mg II K line profiles along the slit between $y= -254 \arcsec$ and $y= -244 \arcsec$ where the slit intersects with the surge material. Note that the imaging and spectral snapshots are not concurrent, as by the time the third phase of the propagating oscillation is visible in the upper part of the flux tube, there is no observable emission at the slit position. The mean Doppler velocity detected in the surge material is positive and increases with time from $\sim 50$ km s$^{-1}$ to over 100 km s$^{-1}$ when the bulk of the surge material passes the spectrograph slit. Immediately following the reconnection event the torsional motion in a counter-clockwise direction is first detected. The centres of both Si IV and Mg II K line profiles show positive gradients with respect to distance along the slit, i.e. $\mathrm{d}v_\mathrm{D}/\mathrm{d}y > 0$. Approximately 40 s later, the direction of rotation reverses and $\mathrm{d}v_\mathrm{D}/\mathrm{d}y < 0$. Finally, during the last phase when the surge material remains visible in the spectral data the sense of rotation reverses once more and the line profile gradient becomes positive again. During all three time intervals, the position of the line centres is not a perfectly monotonic function of $y$. This suggests the presence of further small-scale motions within the flux tube in addition to the large-scale torsional motion.

Both Si IV and Mg II K line profiles are fitted with a single peak Gaussian, as the central reversal in the Mg II K line core is not present in most of the features observed off-limb. Figure \ref{fig:profile_evolution} shows the evolution of the line profile parameters at three locations along the spectrograph slit, marked A, B and C in Fig. \ref{fig:flux_rotation}, for a time interval of 180 s when the surge material is observable in the spectral data. Locations A and C correspond to the two edges of the flux tube and B corresponds to the centre of the flux tube. The peak Si IV and Mg II K line intensities gradually decrease with time as the bulk of the surge material passes the slit, with the exception of the top edge of the flux tube near location A, where the surge plasma is more sparse. The torsional oscillation is clearly visible in the temporal evolution of the Doppler velocity as an anti-phase oscillation at the opposite edges of the flux tube. The Si IV and Mg II K spectral line widths first rapidly increase, which is then followed by more gradual decrease. This suggests that the plasma is gradually cooling except for a short period of impulsive heating immediately after the reconnection event. The evolution of all three profile parameters becomes less clear after the initial 150 s of the IRIS observing sequence, as most of the surge material has by then moved past the slit and is only visible in the upper part of the IRIS SJI field of view.

The Doppler velocity oscillation at opposite edges of the flux tube (locations A and C) is obtained from the Si IV spectra by removing the linear trend from the Doppler velocity time series. The linear trend is due to the line-of-sight component of the surge bulk motion. We fit a damped sine function of the form $\delta v_{\mathrm{D}}(t) = v_0 \exp(-t/\tau) \sin(\frac{2 \pi t}{P} + \phi_{\mathrm{A,C}})$. The amplitude $v_0$, period $P$, damping scaling time $\tau$ and phase $\phi$ are free parameters fitted using weighted least squares to both A and C time series simultaneously subject to the constraint that $\phi_{\mathrm{A}} - \phi_{\mathrm{C}} =  \pi$. The weights are the inverse of the mean relative residuals of the individual Gaussian line profile fits at each time step, which are a measure of goodness-of-fit of a simple Gaussian function for the line profile. We find $v_0 = 41$ km s$^{-1}$, $P = 89$ s and $\tau = 136$ s.  It should be noted that the phase difference between the two time series is in fact expected to be slightly less than as $180^{\circ}$ as the spectrograph slit is not perpendicular to the flux tube axis but intersects it at an angle. As the torsional oscillation is propagating, we do not measure at a fixed time the same phase of the oscillation in both locations.

We estimate the speed of the torsional wave propagation to be 170 km s$^{-1}$ in the plane of the sky using time-distance plots along the axis of the surge flux tube. Assuming the angle between the surge propagation and the line of sight to be 45$^\circ$, the speed of the bulk plasma flow in the plane of the sky is the same as the speed along the line of sight, which we estimate to be 70 km s$^{-1}$ from the mean Doppler shift. This leads to the projected net phase speed of the oscillation of 100 km s$^{-1}$ in the plane of the sky resulting in the total phase speed of 140 km s$^{-1}$. This is consistent with expected Alfv{\'e}n speed of $\sim$ 180 km s$^{-1}$ in a comparable structure given by $v_{\mathrm{A}} = B/\sqrt{\mu_0 \rho}$ assuming the plasma density of prominence material of $10^{-10}$ kg m$^{-3}$ and a magnetic field strength of 20 G, i.e. typical values for prominences \citep[e.g.][]{mackay_physics_2010}. 

\section{Discussion and conclusions}

The observed anti-phase oscillations of Doppler velocity perturbations are subject to attenuation with a typical scaling time measured to be 136 s. Torsional oscillations occurring in a non-homogeneous plasma dissipate their energy via phase mixing \citep{Luo2002,soler_energy_2019}. Phase mixing is caused by variations in the plasma quantities in a direction perpendicular to the magnetic field leading to variations in the local Alfv{\'e}n speed \citep{heyvaerts_coronal_1983}. This results in Alfv{\'e}n waves on different magnetic surfaces propagating with different speeds and becoming increasingly out of phase leading to enhanced dissipation. However, for this mechanism to be observed, it would be necessary to track the wave amplitude of the propagating wave, which is not possible in a fixed-slit observation. Instead, we measure the amplitude profile of the torsional wave train. 

Accumulation of magnetic twist is common both in lower solar atmosphere in the form of small-scale twisted magnetic elements \citep{pontieu_prevalence_2014} and at coronal heights in prominence flux ropes \citep{mackay_physics_2010}. The mechanism of reconnection-induced twist release and associated generation of torsional oscillations is easier to resolve in large-scale structures, however, occurrence of such process can be expected to be ubiquitous on smaller scales.
The magnetic twist originates from twisting and braiding of the footpoints of magnetic structures from magnetoconvective vortex flows \citep{shelyag_vorticity_2011}, or from magnetic flux ropes emerging from the solar interior into the solar atmosphere \citep{archontis_magnetic_2012}. Magnetic reconnection is one of the physical processes responsible for the transfer of magnetic helicity from the lower solar atmosphere into the corona. 
  
We present the first spatially and temporally resolved observation of a torsional Alfv{\'e}n wave in a large-scale structure at coronal heights, detecting clear anti-phase oscillation of line of sight velocities at the opposite edges of the flux tube. We have shown that the magnetic reconnection leads to the generation of torsional Alfv{\'e}n waves. This suggests that the omnipresent small-scale reconnection events occurring in the lowermost layers of the solar atmosphere involving twisted magnetic elements are capable of Alfv{\'e}n wave excitation on global scales.  

\begin{acknowledgements}
This research was supported by the Research Council of Norway through its Centres of Excellence scheme, project no. 262622. E.V. acknowledges financial support from the UK STFC on the Warwick STFC Consolidated Grant ST/L000733/I. C.F. acknowledges funding from CNES. The SDO/AIA data are available courtesy of NASA/SDO and the AIA science team. 1.5 SDO/AIA data are available from AIA cutout service web page http://www.lmsal.com/get\textunderscore aia\textunderscore data/. IRIS is a NASA small explorer mission developed and operated by LMSAL with mission operations executed at NASA Ames Research center and major contributions to downlink communications funded by the Norwegian Space Center (NSC, Norway) through an ESA PRODEX contract. IRIS data are available from mission web page http://iris.lmsal.com/search.
\end{acknowledgements}

\bibliography{torsional}

\end{document}